\documentstyle[12pt,aasms4]{article}

\def\be{\begin{equation}}
\def\ee{\end{equation}}
\def\beq{\begin{eqnarray}}
\def\eeq{\end{eqnarray}}

\def\part{\partial}
\def\nn{\nonumber}

\def\la{\lambda}
\def\f{{\bf f}}
\def\F{{\bf F}}
\def\k{{\bf  k}}

\def\a{{\bf a}}
\def\A{{\bf A}}

\def\etal{et al.~}
\def\b{{\bf b}}
\def\B{{\bf B}}

\def\v{{\bf v}}

\def\V{{\bf V}}
\def\OV{\overline{\V}}
\def\OB{\overline{\B}}

\def\OA{\overline{\A}}

\begin{document}
\baselineskip=24pt
\begin{center}
{\Large\bf Conservation of Magnetic Helicity and Its Constraint on
$\alpha$-Effect of Dynamo Theory}
\bigbreak
{\large\bf by}
\medbreak
{{\large\bf Hongsong Chou and George B. Field} \\
{\it Harvard-Smithsonian Center for Astrophysics, Cambridge, MA 02138,
U.S.A. \\ \{hchou, gfield\}@cfa.harvard.edu}}
\end{center}
\begin{abstract}
Dynamical studies of MHD turbulence on the one hand, and arguments
based upon magnetic helicity on the other, have yielded seemingly contradictory
estimates for the $\alpha$ parameter in turbulent dynamo theory. Here
we show, with direct numerical simulation of three-dimensional
magnetohydrodynamic turbulence with a mean magnetic field, $\OB$, that
the constraint on the dynamo $\alpha$-effect set by the magnetic
helicity is time-dependent. A time-scale $t_c$ is introduced such that
for $t<t_c$, the $\alpha$-coefficient calculated from the simulation is
close to the result of Pouquet \etal and Field \etal,
$-\frac{\tau_{cor}}{3}( \langle \v \cdot \nabla \times \v \rangle  -
\langle \b \cdot \nabla \times \b \rangle )$; for $t>t_c$, the
classical result of the $\alpha$-coefficient given by the Mean-Field
Electrodynamics is reduced by a factor of $1/\left( {R_m
|\OB|^2/v_{rms}^2} \right)$, as argued by Gruzinov \& Diamond, Seehafer
and Cattaneo \& Hughes. Here, $R_m$ is the magnetic Reynolds number,
$v_{rms}$ the rms velocity of the turbulence, $\tau_{cor}$ the
correlation time of the turbulence, and $\overline B$ is in velocity
unit. The applicability of and connection between different models of
dynamo theory are also discussed.
\end{abstract}

\section{Introduction}
The generation and amplification of magnetic field in many astrophysical
systems are often attributed to the turbulent dynamo effect. Since the
seminal papers by Parker (1955) and Steenbeck, Krause \& R\"adler
(1966), a whole body of theory, namely, {\it Mean Field
Electrodynamics} (MFE) has been developed to explain the dynamics of
magnetic field generation by helical turbulence in a
conducting fluid (Moffatt 1978, or Krause \& R\"adler 1980). In such a
fluid, the velocity field, $\v$, stretches the magnetic field, $\b$, in
such a way that the correlation of $\v$ and $\b$ results in an
electromotive force, $ \langle \v \times \b \rangle $ that amplifies
the mean (large-scale) magnetic field, $\OB$, through the relation
\be
 \langle \v \times \b \rangle  = \alpha \OB.
\ee
Here, the coefficient $\alpha$ represents the so-called $\alpha$-effect
(Moffatt 1978); it is calculated in MFE as $\alpha_{MFE} \sim -\frac{\tau_{cor}}{3}
\langle \v \cdot \nabla \times \v \rangle $, where $\tau_{cor}$ is the
correlation time of the turbulence. MFE is a kinematic theory, in that
the velocity field is prescribed and no back reaction of the magnetic
field on the velocity field is considered. Therefore, its applicability
to circumstances where the velocity field is affected by the growing
magnetic field is questionable. Several authors have extended MFE to include the
quenching of the $\alpha$-effect due to the back reaction
of the magnetic field. By numerically solving the spectral MHD equations
using a closure method known as the EDQNM (Eddy-Damped Quasi-Normal
Markovian) approximation, Pouquet, Frisch \& L\'eorat (1976, hereafter PFL) find that
$\alpha=\int_k \alpha_k dk$, where $\alpha_k$ is determined by, in
Fourier space, the difference between the spectra of the kinetic
helicity correlation function, $\langle \v \cdot \nabla \times \v
\rangle$, and the current helicity correlation function, $\langle \b
\cdot \nabla \times \b \rangle$. A similar result was found by Field,
Blackman \& Chou (1999, hereafter FBC), who consider the back reaction of magnetic field by
treating $\v$ and $\b$ on an equal footing, and give a model in which the
$\alpha$-coefficient (for later purposes, we call it $\alpha_1$) can be
expressed as  
\be
\alpha_1 = -\frac{\tau_{cor}}{3}\left( \langle \v \cdot \nabla \times \v \rangle
- \langle \b \cdot \nabla \times \b \rangle \right)
\ee
for relatively small values of $\left({\overline B}/{v_{rms}}\right)^2$.

However, the nonlinear nature of the problem introduces so much
difficulty that the effect of the back reaction is still under
debate. One of the objections to the application of $\alpha_1$ to the
galactic dynamo has its root in the problem of large magnetic Reynolds number and
small-scale fields. Vainshtein \& Cattaneo (1992, see also Cattaneo \&
Vainshtein 1991) argue that for systems with large $R_m$, $\alpha$ is
reduced by a factor of $R_m$ from its kinematic value, i.e.,
\be
\alpha \propto \frac{1}{R_m},
\ee
as the small-scale magnetic field grows quickly to turn off the generation of magnetic
flux at large scales. These authors applied the conservation of the square
vector potential of 2D ideal MHD. The conservation of magnetic helicity
of 3D MHD in steady state was later studied by Seehafer (1994, 1995) who provided
another model of the dynamo $\alpha$-effect in which the
$\alpha$-coefficient (for later purposes, we call it $\alpha_2$) is
given by 
\be
\alpha_2 = -\frac{\la \langle \b \cdot \nabla \times \b
\rangle}{{\overline B}^2}.
\ee
Because $R_m = Lv_{rms}/\la$, $\alpha_2 \propto 1/R_m$ also. Because $R_m
\gg 1$ in astrophysical systems such as the Galaxy, both relation (3)
and relation (4) suggest strong suppression of the dynamo
$\alpha$-effect.  

Another model of the dynamo $\alpha$-effect was introduced by Gruzinov \&
Diamond (1994, 1995, 1996) by studying how the conservation of magnetic
helicity affects the result like relation (2). They realized that because Ohmic
dissipation of the current helicity $\langle \b \cdot \nabla \times \b
\rangle$ in (2) changes the magnetic helicity, the dynamics of the
latter will affect $\alpha_1$, which should be modified to (for later
purposes, we call it $\alpha_3$)
\be
\alpha_3 =  \frac{\alpha_{MFE}}{1+R_m ({\overline B}/v_{rms})^2}.
\ee
Since $R_m$ is usually very large in astrophysics, this again implies
that $\alpha$, in steady state, is far smaller than its classical
value, so that it is too small to be important for the generation of
large scale magnetic fields. Note that in the limit of $R_m \gg 1$ and
$v_{rms} \sim b_{rms}$, $\alpha_3$ is reduced to $\alpha_2$. 

$\alpha_2$ and $\alpha_3$ are different from $\alpha_1$ in the
following two aspects: first, $\alpha_1$ does not explicitly show strong
suppression of the $\alpha$-effect for large $R_m$; second, neither of
the dynamical studies of PFL and FBC, which give $\alpha_1$, explicitly
considered the conservation of magnetic helicity that led to $\alpha_2$
and $\alpha_3$. Under what conditions the magnetic helicity
constraint, which is essential in the models of $\alpha_2$ and
$\alpha_3$, enters the derivation of $\alpha_1$ is still an open
question. In other such words, if such magnetic helicity
constraint can be relaxed in systems of large $R_m$, which of these three
models remains valid? To provide insight into this problem, in the
following we study the dynamics of magnetic helicity using the magnetic
helicity conservation equations, and the time dependence of $\alpha$
using a numerical simulation of MHD turbulence. We find that both the
magnetic helicity development and the $\alpha$-coefficient are time
dependent. We find that the classical MFE result is valid up to a
critical time that we calculate, and the magnetic helicity constrained
result is valid thereafter. Which value to apply therefore depends on
the circumstances.    

This paper is presented in the following structure: in Section 2 we
provide a model for the time-dependence of the magnetic helicity
development, and introduce a critical time, $t_c$, to separate the two
important stages of the magnetic helicity evolution; in Section 3.1 we
present our numerical model that is used to study the time-dependence
of the dynamo $\alpha$-effect; the numerical results are given in
Section 3.2; in Section 4 we discuss the implications of our analytic
model and the numerical results, which are applied to the Galactic
dynamo; conclusions are given in Section 5.

\section{The Time Dependence of Magnetic Helicity Dynamics}

We start from studying the dynamics of the magnetic helicity and its
constraint on the dynamo $\alpha$-effect. For a 3D incompressible MHD
system, we separate the vector potential $\A$ into a large-scale part
$\OA$ and a small-scale part $\a$. Similarly, we write the magnetic
field as $\B = \OB + \b$ and the  velocity field as $\V = \OV + \v$. By
un-curling the induction equation for $\B$,
\be
\frac{\part \B}{\part t} = \nabla \times \left( \V \times \B \right) +
\la \nabla^2 \B,
\ee
we have the equation for $\A$,
\be
\frac{\part \A}{\part t} = \V \times \B + \la \nabla^2 \A - \nabla \psi.
\ee
Here $\la$ is the magnetic diffusivity and $\psi$ is the scalar
potential. Dotting (6) with $\a$ and (7) with $\b$ and summing the
resulting equations together, we have the equation for the ensemble
average of the small-scale magnetic helicity
\beq
\frac{\part}{\part t}  \langle \a \cdot \b \rangle  &=& -2\la  \langle
\b \cdot \nabla \times \b \rangle  - 2 \langle \v \times \b \rangle
\cdot \OB \nn\\ &+&  \langle  \nabla \cdot \left (-\b \psi^{\prime} -
\a \times (\v \times \b + \OV \times \b + \v \times \OB ) \right)
\rangle ,
\eeq
where $\psi^{\prime}$ is the fluctuating component of the scalar
potential. The third term in (8) comes from the $\v \times \OB$ term in
both (6) and (7), which represents the interaction between the
small-scale velocity field $\v$ and the large-scale magnetic field
$\OB$. The physics of this term can be explained as follows. The line
stretching, twisting and folding of $\OB$ will produce $\b$ and $\a$,
therefore affect the generation and diffusion of $ \langle \a \cdot \b
\rangle $ within a volume $V$. The divergence term shows that flux of
magnetic helicity of certain sign can escape from the system through
open boundaries (Blackman \& Field, 2000). The equation for the large-scale
magnetic helicity, $\OA \cdot \OB$, can be also derived. The equations
for $\OB$ and $\OA$ are
\be
\frac{\part \OB}{\part t} = \nabla \times \left( \OV \times \OB \right)
+ \nabla \times  \langle \v \times \b \rangle  + \la \nabla^2 \OB,
\ee
\be
\frac{\part \OA}{\part t} = \OV \times \OB +  \langle \v \times \b \rangle  + \la
\nabla^2 \OA - \nabla {\overline \psi}.
\ee
Following similar procedures that led us to get (8),  we have
\beq
\frac{\part}{\part t}  (\OA \cdot \OB)  &=& -2\la \OB \cdot \nabla \times
\OB + 2  \langle \v \times \b \rangle  \cdot \OB \nn\\ 
&+&  \langle \nabla \cdot \left( - \OB {\overline \psi} - \OA \times (\OV
\times \OB +  \langle \v \times \b \rangle ) \right)  \rangle .
\eeq
Note that the third term in (11) has the opposite sign of the third term
in (8), showing that the electromotive force $ \langle \v \times \b
\rangle $ generates  $\OA \cdot \OB$ of the opposite sign of $\langle
\a \cdot \b \rangle$. If we define the following two flux terms
\be
{\bf f} =  \langle \b \psi^{\prime} + \a \times (\v \times \b + \OV \times \b +
\v \times \OB) \rangle ,
\ee
\be
{\bf F} =  \langle \OB {\overline \psi} + \OA \times (\OV \times \OB +
\langle \v \times \b \rangle ) \rangle,
\ee
with (12) and (13), we may re-write (8) and (11) in the limit of $\la
\rightarrow 0$ in the form
\be
D_t  \langle \a \cdot \b \rangle  \equiv \frac{\part}{\part t}  \langle \a \cdot \b \rangle  + \nabla
\cdot {\bf f} = -2{\cal E} \cdot \OB,
\ee
\be
D_t  (\OA \cdot \OB)  \equiv \frac{\part}{\part t}  \OA \cdot \OB  + \nabla
\cdot {\bf F} = 2{\cal E} \cdot \OB,
\ee
where ${\cal E} =  \langle \v \times \b \rangle $ is the electromotive force. The
conservation of total magnetic helicity, $H_m =  \langle \a \cdot \b \rangle  +  \OA \cdot
\OB $, immediately follows from the above two equations and reads
\be
{\cal D}_t H_m \equiv D_t  \langle \a \cdot \b \rangle  + D_t (\OA \cdot \OB)  = 0,
\ee
which holds for ideal MHD. 

Seehafer (1994, 1995) related the dynamics of small-scale magnetic
helicity, i.e., equation (8), to the quenching of the dynamo
$\alpha$-effect. By assuming stationarity of the MHD turbulence in a
closed system, he argued that $\part_t  \langle \a
\cdot \b \rangle = 0$ and $\nabla \cdot \f = \nabla \cdot \F = 0$;
therefore, by neglecting the time-dependent term and the boundary
term in equation (8) and applying relation (1), he obtained relation
(4) for $\alpha_2$. The same assumptions about the stationarity and the
closedness of the system were made by Gruzinov \& Diamond (1994, 1995,
1996), where they provide a modified $\alpha$-coefficient in the form of
$\alpha_3$. $\alpha_3$ can be regarded as an interpolation between the
MFE result $\alpha_{MFE}$ and the quenching result $\alpha =
\frac{\alpha_{MFE}}{R_m ({\overline B}/v_{rms})^2}$ (see also Cattaneo
\& Vainshtein, 1991 and Vainshtein \& Cattaneo, 1992), thus it is
supposed to be valid for both large and small $R_m$. The numerical
simulation by  Cattaneo \& Hughes (1996) with a particular $R_m = 100$
and different values of $\OB$ supports this result.   

$\alpha_2$ and $\alpha_3$ are different from $\alpha_1$ in that both of
these two models show strong suppression of $\alpha$ for large
$R_m$. The large magnetic Reynolds number $R_m$ in real astrophysical systems
constrains the dynamo $\alpha$-effect to such a degree that, according
to these two models, $\alpha$ would be too small to be important. However,
from the above derivation of both $\alpha_2$ and $\alpha_3$, one can
see that three conditions must be met in order for such constraint to
be effective in real astrophysical systems:
\begin{itemize}
\item a large magnetic Reynolds number,
\item the system is in stationary state, and
\item the system is closed, i.e, no net flux of magnetic helicity flowing through
the system.
\end{itemize}
Because the numerical simulation of Cattaneo \& Hughes (1996) satisfies
all these conditions, it is interesting that their numerical results
confirm $\alpha_3$ in relation (5).  

For real astrophysical systems, these three conditions may not be
satisfied simultaneously. For example, Blackman \& Field (2000) have
argued that most astrophysical objects are open systems, and
magnetic helicity can flow through the boundaries. For systems where $\OB$ cannot
be assumed constant, Bhattacharjee \& Yuan (1995) suggested that
$\alpha \rightarrow \frac{1}{{\overline B}^2} \nabla \cdot \left(
\kappa^2 \nabla\frac{{\overline {\bf J}} \cdot \OB}{{\overline B}^2}
\right)$ for $R_m \rightarrow \infty$, where ${\overline {\bf J}} =
\nabla \times \OB$ and $\kappa^2$ is a positive functional of the
statistical properties of the MHD turbulence. 

Another assumption made in deriving $\alpha_2$ and $\alpha_3$, the
stationarity of the MHD turbulence, may also not be valid when the 
$\alpha$-coefficient in relations (4) or (5) is applied  to real
astrophysical systems. One scenario that we can imagine is that when
impulsive, transient phenomena such as solar flares or supernova
explosions happen, steady astrophysical systems will be
disturbed. This means that, to correctly understand the relation
between the dynamics of magnetic helicity and the dynamo
$\alpha$-effect and its application to real astrophysical systems, one
should study not only the stationary state where the magnetic helicity
has been built up in the system, but also the non-stationary state when
there is net magnetic helicity being built up in or thrown away from
the system. During such non-stationary state of the magnetic helicity
development, the velocity field has the freedom to stay independent
from the growing magnetic field, and if so, the derivation of
$\alpha_2$ from just the induction equation may not give a complete
picture for the dynamo $\alpha$-effect in real astrophysical systems
(Kulsrud 1999). On the other hand, both PFL and FBC, who give
$\alpha_1$ in dynamical studies, incorporate not only the momentum
equation but also the induction equation into their models. Because
$\alpha_3$ can be derived by relating the $\langle \b \cdot \nabla
\times \b \rangle$ terms in relations (4) and (2), it is expected that
$\alpha_3$ should include not only the dynamical considerations of PFL
and FBC but also the magnetic helicity constraint of Seehafer. However,
in order to derive $\alpha_3$ in this way, one has to assume that
$\alpha_1$ is equivalent to $\alpha_2$ in order to get $\alpha_3$ in
the form of relation (5). This means that the assumptions of closedness,
stationarity and large $R_m$ must be satisfied in the model of
$\alpha_3$.

If the stationarity assumption is removed,
the magnetic helicity constraint shown in the models of $\alpha_2$ and
$\alpha_3$ can then be relaxed, even though in these models the
magnetic Reynolds number is large and the system is closed. And if the
magnetic helicity constraint is relaxed, the model of $\alpha_1$ may
give a complete picture of the dynamo $\alpha$-effect. If the relaxation
of the magnetic helicity constraint lasts for a limited period of time,
the model of $\alpha_1$ must be extended so that the magnetic helicity
conservation is taken into account by introducing $\alpha_2$ and
$\alpha_3$. Note that $\alpha_2 \sim \alpha_3$ for large $R_m$, and we
are only interested in large $R_m$ (or small $\la$, as in astrophysics)
case.  

To test our hypothesis of the non-stationary state of a 3D MHD system,
we consider a closed system and neglect the boundary effects due to
$\nabla \cdot \f$ and $\nabla \cdot \F$. The resulting equations for
the small and large scale magnetic helicity can be written as 
\be
\frac{\part}{\part t}  \langle \a \cdot \b \rangle  = -2\la  \langle
\b \cdot \nabla \times \b \rangle  - 2 \alpha {\overline B}^2,
\ee
and
\be
\frac{\part}{\part t}  (\OA \cdot \OB)  = -2\la \OB \cdot
\nabla \times \OB + 2 \alpha {\overline B}^2,
\ee
where relation (1) is used. The interpretation of equations (17) and
(18) is that: the production of positive/negative large-scale magnetic
helicity, $\OA \cdot \OB $, by the dynamo $\alpha$-effect is due to the
production of negative/positive small-scale magnetic helicity, $
\langle \a \cdot \b \rangle $ and to dissipation, $\la  \langle \OB
\cdot \nabla \times \OB \rangle $. For systems with $\la \rightarrow
0$, the total magnetic helicity, $\OA \cdot \OB +  \langle \a \cdot \b
\rangle $, is conserved, and dynamo $\alpha$-effect can be understood
as a {\it pumping effect} that transfers small-scale magnetic helicity
to large scales without generating any total magnetic helicity in the
(closed) system.  To illustrate such a pumping effect, consider the
case that at certain time the small-scale magnetic helicity is zero, $
\langle \a \cdot \b \rangle =0$. A dynamo $\alpha$-effect with positive
$\alpha$ coefficient will pump positive magnetic helicity from small
scales to large scales, and leave negative magnetic helicity at small
scales, so that after some critical time $t_c$, $ \langle \a \cdot \b
\rangle  < 0$, $\OA \cdot \OB > 0$, and $ \langle \a \cdot \b \rangle +
\OA \cdot \OB =0$. 

To quantify these effects, consider the case in (17) that
$\la$ is very small so that the diffusion term can be
neglected. If at $t=0$, $\langle \a \cdot \b \rangle = 0$, the integral
of the resulting equation  gives\footnote{Note that here we are
assuming that $\OB$ is a independent of time, as in our numerical
simulation.} 
\be
\langle \a \cdot \b \rangle (t) = -2{\overline B}^2 \int_0^t
\alpha(\tau) d\tau.
\ee
For small times, when the constraint of magnetic helicity on the dynamo
effect is not yet effective, we assume that
\be
\alpha(t)=\alpha_1,
\ee
the dynamical value calculated by the model of $\alpha_1$, so
\be
\langle \a \cdot \b \rangle (t) \cong -2 {\overline B}^2 \alpha_1
t.
\ee
This should be valid for small values of $t$. However, as $t$
increases, $\langle \a \cdot \b \rangle (t)$ approaches the maximum
that can be associated with the given small-scale magnetic energy. It
can be shown from the realizability condition (Moffatt 1978) that
\be
|\langle \a \cdot \b \rangle(t)| \leq 2Lb_{rms}^2,
\ee
where $L$ is the outer scale of the turbulence, and $b_{rms}$ is the
rms small-scale field strength. In light of this, (20) is valid only up
to a critical time $t_c$ given by
\be
t_c \cong \frac{Lb_{rms}^2}{\alpha_1 {\overline B}^2}.
\ee
Since $\alpha_1 \sim {\cal O}(v_{rms})$, where $v_{rms}$ is the rms
velocity, and $L/v_{rms}=t_{eddy}$, the eddy turnover time,
\be
t_c \cong t_{eddy}\left( \frac{b_{rms}}{\overline B} \right)^2.
\ee
A more realistic estimate of $t_c$ should include a correction factor,
which modifies the above formula to $t_c \cong C t_{eddy}\left
( \frac{b_{rms}}{\overline B} \right)^2$. The correction factor $C$
depends on the ratio $v_{rms}/\alpha_1$ and the ratio
$|\langle \a \cdot \b \rangle|/2Lb_{rms}^2$. $C$ can be estimated from
numerical simulation results, as we will do in the next section.

When $t>t_c$, because there is no further small-scale magnetic helicity to draw
on, $\part_t \langle \a \cdot \b \rangle \rightarrow 0$, and from (17)
we have
\be
\alpha {\overline B}^2 = -\la \langle \b \cdot \nabla \times \b
\rangle,
\ee
which implies that $\alpha$ no longer equals $\alpha_1$, but rather
is determined by the fact that the only available source for the
large-scale magnetic helicity being pumped by $\alpha$ is that which is
being dissipated at small scales with appropriate sign. That is,
$\alpha$ equals $\alpha_2$, hence is suppressed by the large $R_m$ (or
small $\la$) for the stage $t > t_c$. To estimate the value of the
suppressed $\alpha$, $\alpha_{sp}$, we replace the current
helicity of the small-scale magnetic field, $\langle \b \cdot \nabla
\times \b \rangle$, with the value corresponding to a maximally helical
small-scale field (which we assumed to result from (20) at $t=t_c$),
which is  
\be
|\langle \b \cdot \nabla \times \b \rangle| \cong L^{-1}b_{rms}^2.
\ee
Hence
\be
|\alpha_{sp}| \sim \frac{\la}{L}\left( \frac{b_{rms}}{\overline
B}\right)^2.
\ee
Since $|\alpha_1| \sim v_{rms}$, and $R_m = Lv_{rms}/\la$, this can
be written as
\be
|\alpha_{sp}| \sim \alpha_1 \frac{1}{R_m} \left
( \frac{b_{rms}}{\overline B} \right)^2.
\ee
This formula is consistent with $\alpha_3$ when $R_m$ is large, and is 
identical to $\alpha_2$, the model by Seehafer (1994, 1995). However,
notice that unlike previous derivations, ours depends critically on
time; so 
\be
\frac{\alpha}{\alpha_1} = \left\{ \begin{array}{ll} 1 &
\mbox{$t<t_c=t_{eddy}\left(\frac{b_{rms}}{\overline B}\right)^2$,} \\
\frac{1}{R_m}\left( \frac{b_{rms}}{\overline B}\right)^2 &
\mbox{$t>t_c$.} \end{array} \right.
\ee

Note that neither $\alpha_2$ nor $\alpha_3$ appears in relation (29)
because they should be valid only for $t>t_c$, when the magnetic helicity
constraint takes effect.  

In the above discussion, we made the following assumptions that require
further justification: (1) the system has an amount of small-scale
magnetic helicity that can be estimated as $Lb_{rms}^2$, where $L$ is the
scale where the magnetic helicity is concentrated; (2) current helicity
can be estimated as $b_{rms}^2/L$; (3) $v_{rms}^2 \sim b_{rms}^2$. To
test these assumptions and calculate $t_c$ as a function of
$t_{eddy}b_{rms}^2/{\overline B}^2$, we apply direct numerical simulation to a 3D
incompressible MHD system with periodic boundary conditions. The
numerical model and simulation results are presented in next
section. 

\section{Numerical Model and Simulation Results}
\subsection{Numerical Model}
Under an external force $\cal F$, the undimensionalized incompressible
MHD equations can be written as (with Einstein summation convention)
\be
\left( \partial_t - \nu \nabla^2 \right) v_i = \partial_j
\left(-p\delta_{ij} - v_iv_j + b_ib_j \right) + {\cal F}_i,
\ee
\be
\left( \partial_t - \la \nabla^2 \right) b_i = \partial_j
\left(v_ib_j - b_iv_j \right),
\ee
\be
\partial_iv_i = \partial_jb_j = 0,
\ee
where $\nu$ and $\la$ are the molecular viscosity and magnetic
diffusivity, respectively. Note that we have written $b$ in units of
$\sqrt{4\pi\rho}$ after dividing both sides of the Navier-Stokes
equation and the induction equation by density $\rho$. If we
use a hat, $\wedge$, to denote discrete Fourier transform, and $\otimes$ to
denote convolution, the above equations in Fourier space are
\be
\left( \part_t + \nu k^2 \right) {\hat v}_j = P_{jl} \left
[ \mbox{i}k_m\left(-{\hat v}_l \otimes {\hat v}_m + {\hat b}_l \otimes
{\hat b}_m \right) + {\hat {\cal F}}_l \right],
\ee
\be
\left( \part_t + \la k^2 \right) {\hat b}_j = P_{jl} \left
[ \mbox{i}k_m\left({\hat v}_l \otimes {\hat b}_m -{\hat v}_m \otimes
{\hat b}_l \right) \right],
\ee
\be
k_m{\hat v}_m = k_j{\hat b}_j = 0.
\ee
Here $\bf P$ is the projection operator defined as $P_{jl} =
\delta_{jl} - \frac{k_j k_l}{k^2}$. In our simulation, we treat the
system as a cube $[0, 2\pi) \times [0, 2\pi) \times [0, 2\pi)$. The
Cartesian coordinate of a grid point can be written as $ x_l =
\frac{2\pi}{N}l, y_m=\frac{2\pi}{N}m, z_n=\frac{2\pi}{N}n,\mbox { for
$l,m,n=0,1,2,...,N-1$}$. A point in Fourier space has coordinates $k_s
= s, k_p=p, k_q=q,\mbox { for $s,p,q=-\frac{N}{2}, -\frac{N-1}{2}, ...,
\frac{N}{2} - 1$}$. Because we assume periodic boundary conditions, the
surface terms $\nabla \cdot f$ and $\nabla \cdot \F$ in equations (14)
and (15) both vanish.

Equations (33), (34) and (35) are numerically solved with the standard
Fourier spectral method. Equations (33) and (34) are treated as ordinary
differential equations for ${\hat v}$ and ${\hat b}$. With the
projection operator {\bf P}, the divergence free condition (35) will be
satisfied for $t>0$ as long as ${\hat v}$ and ${\hat b}$ are divergence
free at $t=0$. All our simulation runs start from divergence free initial
conditions. We employ a second-order Runge-Kutta (RK2) method to advance
equations (33) and (34) in time. We can exploit the advantage of using
RK2 in the following two aspects. First, an integral factor can be
easily introduced with the transform
\be
{\cal U}_m (t) = {\hat v}_m (\k, t) e^{-\nu k^2 t}, {\cal B}_m (t) =
{\hat b}_m (\k, t) e^{-\la k^2 t}.
\ee
Second, aliasing errors can be reduced by introducing positive and
negative random phase shifts at the first and second stages of RK2,
respectively (Machiels \& Deville, 1998). 

The forcing term used in our simulation is the sum of two forcing
functions. In Fourier space, it has the form 
\be
{\hat {\cal F}}(0.5 < |{\bf k}| \leq 1.5) = {\hat {\cal F}}_c + {\hat
{\cal F}}_b.
\ee
That is, the force works only within the shell $S_1: 0.5 < |{\bf k}|
\leq 1.5$. Here ${\hat {\cal F}}_c$ is a forcing term that is similar to, but
not exactly the same as, the one adopted by Chen \etal (1993). It is calculated
by multiplying the velocity components within shell $S_1: 0.5 < |{\bf
k}| \leq 1.5$ by a factor, $\gamma > 1$, so that before a new step of
integration starts, the kinetic energy density within this shell is
reset to $E_1 = 0.24$. Phases of the velocity components within the
shell are not changed. This forcing is equivalent as lengthening the
velocity vectors within shell $S_1$ by a factor $\gamma -1$. Denote the
increment of a velocity vector under force ${\hat {\cal F}}_c$ in
Fourier space as $\delta{\hat {\bf v}} = {\hat {\bf R}} + \mbox{i}{\hat
{\bf I}}$, where ${\hat {\bf R}}, {\hat {\bf I}}$ are the real
and imaginary parts of $\delta{\hat {\bf U}}$. In our simulation, we need to
inject kinetic helicity, $\v \cdot \nabla \times \v$, into the
turbulence. To do this with ${\hat {\cal F}}_c$, we tune the angle
between ${\hat {\bf R}}$ and ${\hat {\bf I}}$ so that they remain
perpendicular to each other. Because kinetic helicity at $\k$ can be
calculated as $H(\k) = 2\k \cdot {\hat {\bf R}} \times {\hat {\bf I}}$,
by doing such ``lengthening'' and ``angle-twisting'', we inject not
only the kinetic energy but also the kinetic helicity into the
turbulence.  

The forcing term ${\hat {\cal F}}_c$ maintains the energy level at the
forcing scale so that the fluctuation in the energy development history
can be small; therefore, the growth stages of both the kinetic energy
and the magnetic energy can be studied carefully and
accurately. However, this force does not introduce random phases into
the velocity field. To be more realistic about the forcing in our
simulation, we also add the forcing term ${\hat {\cal F}}_b$, derived
from the forcing function used by Brandenburg (2000), as a secondary
forcing function to introduce random phases into the velocity
field. ${\hat {\cal F}}_b$ has the form
\be
{\hat {\cal F}}_b(\k) = {\hat {\cal F}}_0 \frac{\k \times (\k \times
{\bf {\hat e}}) - {\mbox i}|\k| (\k \times {\bf {\hat e}})}{2k^2
\sqrt{1-(\k \cdot {\bf {\hat e}}^2)/k^2}}  \cos (\phi (t)).
\ee
Here ${\hat {\cal F}}_0<1$ is a factor adjusted at each time step so
that the kinetic energy density within shell $S_1$ fluctuates within
$\pm5$\% of $E_1$. ${\hat {\bf e}}$ is an arbitrary unit vector in Fourier
space. $\phi(t)$ is a random phase. Note that ${{\hat {\cal F}}_b({\k})}^* =
{\hat {\cal F}}_b(-{\k})$ so it is real, and it is helical in that
${\hat {\cal F}}_b \cdot \nabla \times {\hat {\cal F}}_b = -k {\hat
{\cal F}}_0^2 < 0$, i.e., it has maximum helicity. Because ${\hat {\cal
F}}_b$ is tuned in such way that it only contributes to $\pm5$\% of the
kinetic energy at the forcing scale, ${\hat {\cal F}}_b$ can be considered as a
perturbation to ${\hat {\cal F}}_c$. Therefore, the advantage of using (37) as
the forcing term is three fold: to avoid strong fluctuations of kinetic
and magnetic energy density with time, to introduce random phases to
the velocity field, and to maintain the kinetic helicity at certain
level. 

For each of the simulation runs listed in Table 1, the initial
conditions are set in the following way. The initial velocity field for
each run is a fully developed, pure hydrodynamic, helical, turbulent field,
obtained by applying exactly the same forcing function as that in Chen
\etal (1993) to the Navier-Stokes equation (i.e., no magnetic field
present) with $\nu$ of that run. It is obtained and maintained
helical by applying the  ``angle-twisting'' method mentioned above for
${\hat {\cal F}}_c$.  The initial magnetic field for each run is set up
as a large-scale magnetic field, $\OB$,  along $y$-direction. $\OB$ is
constant in both space and time. A magnetohydrodynamic turbulence
simulation run is then started under the forcing (37) with a set of
initial values of the pure hydrodynamic turbulent velocity field, $\OB$
and $\nu=\la$, some of which are shown in Table 1.  

\subsection{Simulation Results}
To simplify matters, we set $\nu = \la$ in all our simulation
runs, i.e., the magnetic Prandtl number is 1. $\nu(=\la)$ and ${\OB}$
are taken as free parameters of the numerical simulation and set up as
initial conditions. All our simulation runs are performed on a $(64)^3$
spatial resolution. As the MHD turbulence reaches steady state, we
calculate the Reynolds number of the turbulence with the formula
$R_e(=R_m) =  \langle  L  \rangle  v_{rms}/\nu$, where $ \langle  L
\rangle $ is the integral length scale of the system. If $E_{vk}$ is
the kinetic energy spectrum, $L$ is calculated as $ \langle L \rangle
=\sum_k k^{-1}E_{vk} / \sum_k E_{vk}$. At $t=0$, we impose a
$\OB={\overline B}{\hat {\bf y}}$ into a fully developed hydrodynamic
turbulence, and follow the MHD turbulence thereafter. We calculate the
$\alpha$ coefficient through $\alpha =  \langle  \v \times \b  \rangle
_y /{\overline B}$. Table 1 lists all the simulation runs that we
obtained. In Figure 1, we plot the evolution of kinetic energy and magnetic
energy. During the kinematic phase of the development, for the first
few eddy turnover times, magnetic energy density grows as $\sim t^2$,
followed by an exponential growth. The growing magnetic energy density will
impose Lorentz force on the velocity field. The kinematic phase ends when
the Lorentz force is strong enough to significantly change the velocity
field, and magnetic energy growth slows down. There is then a dynamic
phase during which the magnetic energy and the kinetic energy
oscillate around a certain level. During this phase, we can estimate
the rms values of both the velocity field and the magnetic field. They
are calculated using the formulae $v_{rms} = \sqrt{2E_v/3}$ and
$b_{rms} = \sqrt{2E_b/3}$, where $E_v$ and $E_b$ are the kinetic energy
density and magnetic energy density. In Figure 2, we plot the kinetic
and magnetic energy spectra for the case of ${\overline B}=0.1$. At
scales of $k < 5$, the kinetic energy density surpasses the magnetic energy
density, showing that near the outer scales, the turbulence is largely
hydrodynamic in nature. For $k \geq 5$, the kinetic energy density is
smaller than the magnetic energy density by a factor less than
three\footnote{Exact equipartition between the kinetic energy and the
magnetic energy was not found in our simulation. This may be due to the
fact that our numerical code has a relatively low spatial resolution ($64^3$) so that
the MHD turbulence inertial range resolved in our simulation is
not very long. Note that the theoretical implication of the
equipartition (see Blackman \& Field, 2000) between the kinetic energy
and the magnetic energy only applies to the inertial range, not the
dissipation range, whereas the non-equipartition found in Figure 2 and
other works (see Brandenburg 2000, Fig. 11, and Cho \& Vishniac, 2000,
Fig. 7) appears mostly in the dissipation range.}.  We also plot the
spectra of the absolute values of the kinetic helicity spectrum $K_k$,
the current helicity spectrum $C_k$ and the magnetic helicity spectrum
$M_k$ (see the caption of Figure 2 for detailed definition of these
spectra). In our simulation, we always force maximally (negative)
helical flow within shell $S_1: 0.5 < |{\bf k}| \leq 1.5$. This
explains the relation $2E_{vk}(k=1) = |K_k(k=1)|$ in the plot. For
$k>1$, the flow is not maximally helical. Rather, $K_k$ decreases as
$k$ increases in a way similar to $E_{vk}$, as it decays into small
scales (large $k$). The current helicity, $C_k$, is also concentrated
at large scales, $k \sim 1$, and decreases as $k$ increases. In all of
our simulation runs, we find that 90\% of each of the kinetic helicity, the
current helicity and the magnetic helicity are concentrated near the
outer scales of the turbulence, i.e., $k \le 4$. In deriving relations
(23) and (27), we assumed that magnetic helicity and current helicity
are both concentrated near the outer scales of the turbulence, and
approximated the total magnetic helicity and the total current helicity
with $2Lb_{rms}^2$ and $b_{rms}^2/L$, respectively. From our simulation
we find that such assumptions are justified, and adopt $L \sim  \langle
L \rangle $. Our assumption of $v_{rms}^2 \sim b_{rms}^2$ is not valid
for all of the cases. Therefore, when comparing our estimation of the
suppressed dynamo $\alpha$ coefficient (equation (27)) with the result
of numerical works (for example, Cattaneo \& Hughes (1996)), we must
take this factor into account.  

To study the relation between the dynamo $\alpha$-effect and the dynamics
of magnetic helicity, we re-write equation (17) in the form
\be
2  \langle \v \times \b \rangle  \cdot \OB + 2\la  \langle  \b \cdot
\nabla \times \b \rangle  + \frac{\part}{\part t}  \langle \a \cdot \b
\rangle  = 0. 
\ee
We calculated the numerical results of each of the three quantities and
the sum of them, and plotted them in Figure 3. Panel (a) of Figure 3 is
the temporal evolution of the quantity $2  \langle \v \times \b \rangle  \cdot
\OB$. The evolution of this quantity can be separated into two
stages. For $t\leq11$, it first increases from 0 to a peak value of
0.28, then decreases until it changes sign at $t\sim11$. After
$t\sim11$, it oscillates around an averaged value of
$0.0027\pm0.0047$. The second term in equation (39), $2\la  \langle  \b
\cdot \nabla \times \b \rangle$, is plotted in panel (b) of Figure 3
with $\la = 0.0133$. This term represents the dissipation effect. It is
a less dominant effect than the dynamo $\alpha$-effect, which is
represented by the first term in equation (39) and plotted in panel
(a). The third term in equation (39), $\part_t  \langle
\a \cdot \b \rangle$, is plotted in panel (c) of Figure 3. Its temporal
behavior is similar to $2  \langle \v \times \b \rangle  \cdot
\OB$. The sum of all these three quantities should be zero for our
closed system with periodic boundary condition, and this is shown in the
bottom panel of Figure 3.

We calculated the $\alpha$ coefficient with $\alpha(t) = \langle \v
\times \b \rangle  \cdot \OB/{\overline B}^2$ and plotted it in the top
panel of Figure 4. The evolution of the small-scale magnetic helicity
density, $\langle \a \cdot \b \rangle$, is shown in the middle panel of
Figure 4. The initial value of the small-scale magnetic helicity is
assumed to be zero. After the start of simulation, negative small-scale
magnetic helicity is built up by the $\alpha$-effect
with positive $\alpha$-coefficient. The speed of this building-up
process achieves its maximal value when the $\alpha$-coefficient
reaches its peak value. After that, the build up of negative
small-scale magnetic helicity slows down as the $\alpha$-coefficient
decreases. As the positive $\alpha$-coefficient approaches zero, the
second term in equation (39), will dominate the $\alpha$ effect term to
affect the dynamics of magnetic helicity. Panel (b) of Figure 4 and
panel (b) of Figure 3 together show that when the dissipation term
becomes important, the negative magnetic helicity decays. Our estimation
of $\alpha(t<t_c)$ is $0.12\pm0.06$, and that of $\alpha(t>t_c)$ is
$0.008\pm0.011$. This clearly shows that the dynamo $\alpha$-effect is
a time-dependent quantity, and the constraint of magnetic helicity does
not take effect on $\alpha$ until the building-up process of magnetic
helicity is almost finished.

To test how close the estimates by the models of $\alpha_1$, $\alpha_2$
and $\alpha_3$ to the measured $\alpha$-coefficient from our numerical
simulation, we take $R=R_m\left({\overline B}^2/v_{rms}^2 \right)$ as a
variable, and plot the measured $\alpha$ and the estimates of $\alpha$
from different models against this quantity. The results are given in
Figure 5. In panel (a) of Figure 5, we plot $\alpha(t<t_c)$,
$\alpha(t>t_c)$, $\alpha_1$, $\alpha_2$ and $\alpha_3$ vs. $R$. It is
clear that $\alpha(t<t_c)$ is close to $\alpha_1$ for all the values of
$R$ considered in our simulation runs, while $\alpha_2$ and $\alpha_3$
underestimated $\alpha(t<t_c)$ for large values of $R$. For $R<1$,
$\alpha_1$ and $\alpha_2$ give similar results. Panel (a) of this
figure also shows that $\alpha_2$ and $\alpha_3$ give much better
estimates of the $\alpha$-coefficient for $t>t_c$, when the constraint
of magnetic helicity on the dynamo $\alpha$-effect finally
enters. Panel (b) of Figure 5 shows a clear linear correlation between
the measured critical time $t_c$ and our estimate of this quantity
using $t_{eddy}\left(b_{rms}/{\overline B}\right)^2$, which we
discussed previously in the Introduction section.   

\section{Discussion}
The appearance of the electromotive force term in the equations for
small-scale and large-scale magnetic helicity, i.e., (8) and (11), shows
that the dynamo $\alpha$-effect is related to the dynamics of magnetic
helicity. But because the dynamo effect is not only determined by the
induction equation but also the momentum equation of the velocity field,
the external forcing term in the momentum equation will provide extra
degrees of freedom to the dynamo $\alpha$-effect, so that it is not
completely determined by the dynamics of magnetic helicity. Instead, as
we have shown in previous sections, the dynamo $\alpha$-effect is
largely controlled by the velocity field during the stage that
small-scale magnetic helicity of appropriate sign is being pumped into
large scales. At the same time as the small-scale magnetic helicity is
being pumped to large scales at this stage, the small-scale magnetic
helicity of opposite sign will be built up. For closed astrophysical
systems of very small magnetic diffusivity, such pumping process is in
fact controlled by the dynamo $\alpha$-effect, multiplied by
${\overline B}^2$. If the initial large scale field ${\overline B}_0$
is very weak, the critical time $t_c = t_{eddy}\left(b_{rms}/{\overline
B}_0\right)^2$ is initially very long, so the dynamical value of the
$\alpha$-effect, i.e., $\alpha_1$ of PFL and FBC, applies, and
considerable amplification of ${\overline B}$ takes place.   

The Galaxy  provides an interesting example. There, $t_{eddy} \sim
10^7$ years, $b_{rms} \sim 3 \times 10^{-6}$ Gauss, and ${\overline
B}_0$ has been estimated as $10^{-13}$ Gauss (Field 1994). Hence
initially $t_c \sim 10^{22}$ years, and dynamo action is not
significantly affected by the build up of magnetic helicity.

During this period, ${\overline B}$ will increase exponentially due to the
interaction of the $\alpha$ effect and the $\omega$ effect associated with the
differential rotation of the Galaxy. To avoid the complications of the
$\omega$ effect here, we will consider the simpler case of an
${\alpha}^2$ dynamo, whose $e$-folding time is $\tau = \lambda/2\pi \alpha$. With a
wavelength of 1Kpc and an $\alpha = 3\times10^4$cm/sec (Field 1994),
$\tau = 5 \times 10^8$ years, or 50 $t_{eddy}$. Presumably growth will
continue until $\tau$  matches the ever decreasing value of $t_c$. This
will occur when ${\overline B} = b_{rms}/50^2 = 5 \times 10^{-7}$
Gauss, after which further growth will be inhibited by the helicity
constraint. To reach this stage will take 15 growth times, or $7.5
\times 10^9$ years. Thus, during a relatively long period, dynamo
growth can occur unconstrained by magnetic helicity, and during this
period, can approach equipartition  within an order of magnitude.

Bear in mind that this example is oversimplified. In particular, it
does not take into account that helicity  may escape through the
boundaries of the system (Blackman \& Field 2000). In this case, the
boundary terms $\f$ and $\F$ in equations (14) and (15) must be taken
into account, and so one must be cautious about applying conclusions from
simple models like we have discussed to real astrophysical systems.

The constraint of magnetic helicity on the dynamo $\alpha$-effect, as
discussed by Gruzinov \& Diamond (1994, 1995, 1996), Cattaneo \& Hughes
(1996) and Seehafer (1994, 1995), plays important role in controlling
the amount of magnetic helicity pumped by the dynamo $\alpha$-effect
from small scales to large scales. It is found in this paper that such
constraining effect takes place after the magnetic helicity development
reaches a critical time, $t_c$. Before $t_c$, we believe that the
source of magnetic helicity being pumped from small scales to large
scales may not be restricted to the Ohmic dissipation at small
scales. Rather, it may come from the twisting, folding and stretching
of the magnetic field lines by the velocity field at different
scales. Such interactions between $\v$ and $\b$ alter the topology of
the magnetic field in such a way that net magnetic helicity, which
represents the magnetic field line topology (Moffatt \& Tsinober 1992),
can be built up. Another source of magnetic helicity can be from
outside of the system if there are open boundaries. During such
non-stationary stage, induction equation (6) alone cannot give a
complete picture of the dynamo $\alpha$-effect, and the dynamical 
studies by PFL and FBC may provide a valid $\alpha$-coefficient,
$\alpha_1$, which cannot be equated to $\alpha_2$. When the building up
of magnetic helicity approached its upper limit set by the
realizability condition (22), the dynamical pumping effect described by
$\alpha_1$ starts to be constrained by the magnetic helicity
conservation. Eventually, if the system is closed, $\alpha_1$ alone (in
the form of relation (2)) is not enough for a complete picture of the
dynamo $\alpha$-effect, and $\alpha_2$ must be introduced to get
$\alpha_3$. 

Another motivation of our work is the numerical study of magnetic
helicity by Stribling \etal(1994). In their simulation, they find
that the electromotive force is not suppressed for the first
few eddy turnover times in a 3D MHD turbulence with an imposed
moderately strong large-scale magnetic field and an $R_m
\rightarrow \infty$. Our work is an extension of the work by
Stribling \etal in the following aspects: first, we introduced a
critical time to separate the non-suppressed stage from the suppressed
stage of the $\alpha$-effect; second, we numerically studied the
dependence of the $\alpha$-coefficient on the values of $\overline B$,
$b_{rms}$, $\tau_{eddy}$ and other quantities of the MHD turbulence. By
doing so, we argued that the time behavior of the magnetic helicity
dynamics plays an important role in the dynamo $\alpha$-effect; therefore,
one cannot simply ignore the $\part_t$-terms when applying the magnetic
helicity equations, (17) and (18), to real astrophysical systems.  

\section{Conclusion}
We studied the constraint of magnetic helicity on the dynamo
$\alpha$-effect with 3D direct numerical simulation under periodic
boundary conditions. The dynamics of magnetic helicity affects the dynamo
$\alpha$-effect only after the magnetic helicity at small scales is
built up and the magnetic helicity dynamics enters a stationary
state. Such building-up process can be understood as a pumping effect
of the dynamo $\alpha$-effect, and the $\alpha$-coefficient during this
non-stationary pumping stage can be estimated according to the model by
Pouquet \etal (1976) or Field \etal (1999). As the small-scale magnetic
helicity is built up to the level limited by the realizability
condition, the $\alpha$-effect is quenched, as suggested by Gruzinov \&
Diamond (1994, 1995, 1996) and Cattaneo \& Hughes (1996). The
$\alpha$-coefficient during such magnetic helicity constraining stage
can be estimated according to the model by Seehafer (1994, 1995, see also
Blackman \& Field 2000). A critical time, $t_c \sim
t_{eddy}\left(b_{rms}/{\overline B}\right)^2$, is introduced to
separate these two stages.    

\acknowledgments

We benefited from our discussions with E.G. Blackman, A. Brandenburg,
B. Chandran and R. Kulsrud. We thank an anonymous referee for
insightful comments.

\begin{deluxetable}{cccccc}
\footnotesize
\tablecaption{Measurements of various physical quantities for different
simulation runs}
\tablewidth{0pt}
\tablehead{ \colhead{} & \colhead{Run I} & \colhead{Run II} &
\colhead{Run III} & \colhead{Run IV} & \colhead{Run V} }
\startdata
${\overline B}$  & 0.316 & 0.1 & 0.0316 & 0.0316 & 0.224 \nl
$\nu(=\la)$ & 0.0133 & 0.0133 & 0.0071 & 0.0133 & 0.0133 \nl
$v_{rms}$ & $0.473\pm0.006$& $0.498\pm0.008$ &$0.50\pm0.01$ &
$0.487\pm0.005$ & $0.489\pm0.002$ \nl
$b_{rms}$ & $0.35\pm0.04$ & $0.26\pm0.04$ &$0.24\pm0.07$ &
$0.19\pm0.05$ & $0.27\pm0.07$ \nl
$R_m$ & $29.8\pm0.9$ & $31.2\pm1.4$ &$54.9\pm2.1$ & $28.5\pm0.6$ &
$29.6\pm0.9$ \nl
$t_{eddy}$ & $1.6\pm0.1$  & $1.6\pm0.2$  & $1.6\pm0.7$ & $1.7\pm0.4$ &
$1.7\pm0.2$ \nl
$\tau_{cor}$ & $0.64\pm0.08$ & $0.8\pm0.3$ & $0.7\pm0.4$ & $0.5\pm0.2$ &
$1.0\pm0.4$ \nl
$t_c$ & $\sim11$ & $\sim32$ &$\sim85$ &$\sim120$ &$\sim20$ \nl
$\alpha(t<t_c)\tablenotemark{a}$ & $0.12\pm0.06$ & $0.21\pm0.08$ &$0.26\pm0.09$
&$0.24\pm0.26$ & $0.13\pm0.09$ \nl
$\alpha(t>t_c)\tablenotemark{a}$ & $0.008\pm0.011$ & $0.10\pm0.11$ &$0.20\pm0.09$
&$0.17\pm 0.22$ & $0.04\pm0.02$ \nl
$\alpha_1$\tablenotemark{b}  & $0.11\pm0.05$ & $0.23\pm0.05$ &$0.16\pm0.12$
&$0.22\pm0.05$ & $0.16\pm0.0.01$ \nl
$\alpha_2$\tablenotemark{b}   & $0.006\pm0.006$ & $0.11\pm0.07$ &$0.14\pm0.05$
&$0.18\pm0.17$ & $0.01\pm0.01$ \nl
$\alpha_3$\tablenotemark{b}   & $0.007\pm0.003$ & $0.09\pm0.03$ &$0.13\pm0.04$
&$0.19\pm0.06$ & $0.022\pm0.002$ \nl
\enddata
\tablenotetext{a}{Temporal average of the numerical
$\alpha$-coefficient ($=\langle  \v \times \b  \rangle_y /{\overline
B}$) before and after the critical time $t_c$.}
\tablenotetext{b}{Estimates of the $\alpha$-coefficient by three
different models (see equations (37), (38) and (39) for details).}
\end{deluxetable}

\clearpage

\begin{figure}
\plotone{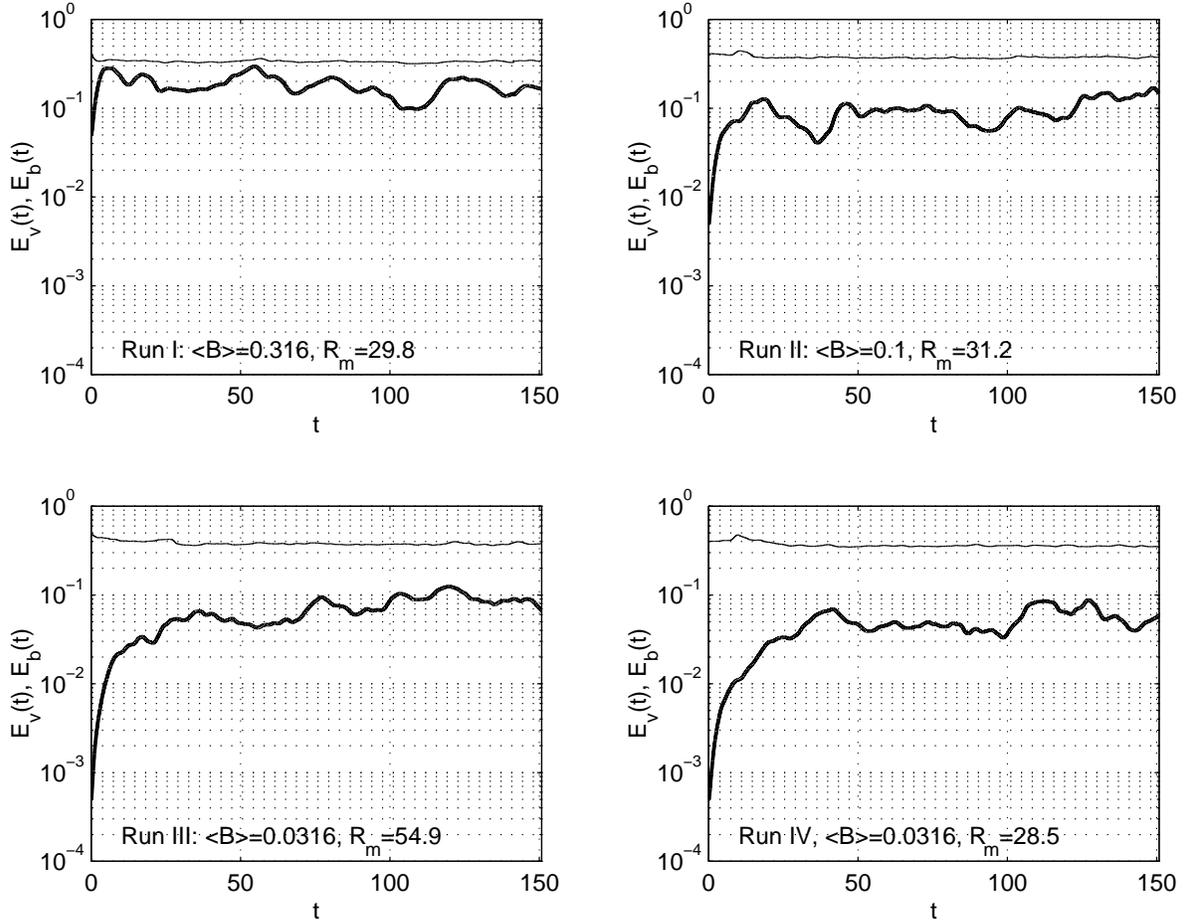}
\caption{Temporal evolution of the kinetic energy density(thin solid line)
and magnetic energy density(thick solid line) for a few run cases.}
\end{figure}

\begin{figure}
\plotone{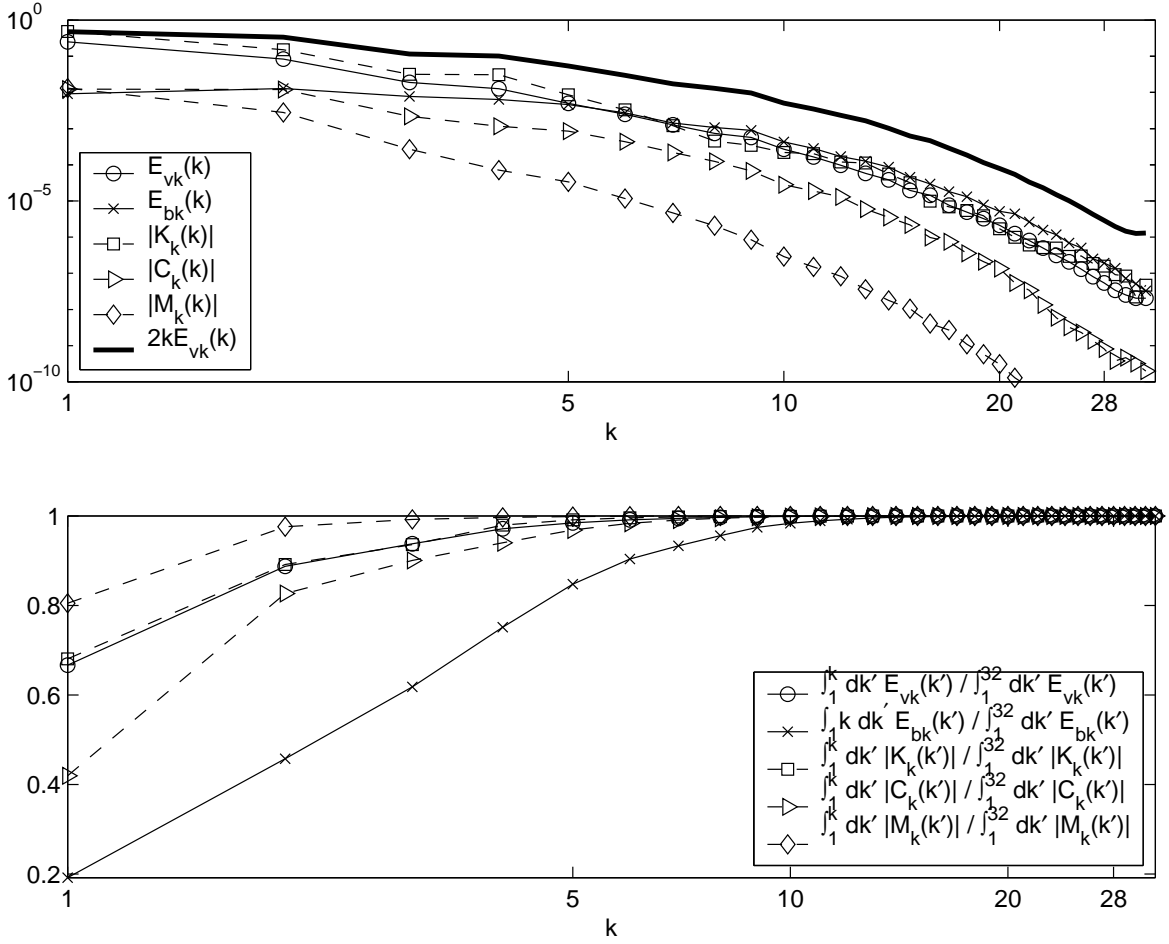}
\caption{Top panel: kinetic energy spectrum,
$E_{vk}(k)=\frac{1}{2}\sum_{k-0.5}^{k+0.5} |{\hat v}(k^{\prime})|^2$;
magnetic energy spectrum, $E_{bk}(k)=\frac{1}{2}\sum_{k-0.5}^{k+0.5}
|{\hat b}(k^{\prime})|^2$; absolute value of the kinetic helicity
spectrum, where $K_{k}(k)=\sum_{k-0.5}^{k+0.5}\mbox{i}\k^{\prime} \cdot ({\hat
\v}(\k^{\prime}) \times {\hat \v}^* (\k^{\prime}))$; absolute value of current helicity
spectrum, where $C_{k}(k)=\sum_{k-0.5}^{k+0.5}\mbox{i}\k^{\prime} \cdot ({\hat
\b}(\k^{\prime}) \times {\hat \b}^* (\k^{\prime}))$; absolute value of
magnetic helicity spectrum, where $M_{k}(k)=\sum_{k-0.5}^{k+0.5}({\hat \a}(\k^{\prime}) \cdot
{\hat \b}^* (\k^{\prime}))$ and $\a$ is the vector potential of $\b$;
kinetic helicity spectrum of maximally helical flow,
$2kE_{vk}(k)$. Relative error for $E_{vk}(k)$ ranges from $\pm5\%$ to
$\pm40\%$ for different $k$'s, with $k=1$ has the smallest
error. Relative error for $E_{bk}(k)$ ranges from $\pm30\%$ to
$\pm50\%$. Bottom panel: cumulative spectra of $E_{vk}, E_{bk},
|K_k|, |C_k|$ and $|M_k|$. Data are collected from Run II, averaged
from time $t=52.5$ to $t=100.5$. For this case, $ \langle L \rangle
=0.80 \pm 0.02$, $v_{rms} = 0.498 \pm 0.008$; therefore Reynolds number
is $R_e = R_m = 29.8 \pm 0.9$.}
\end{figure}

\begin{figure}
\plotone{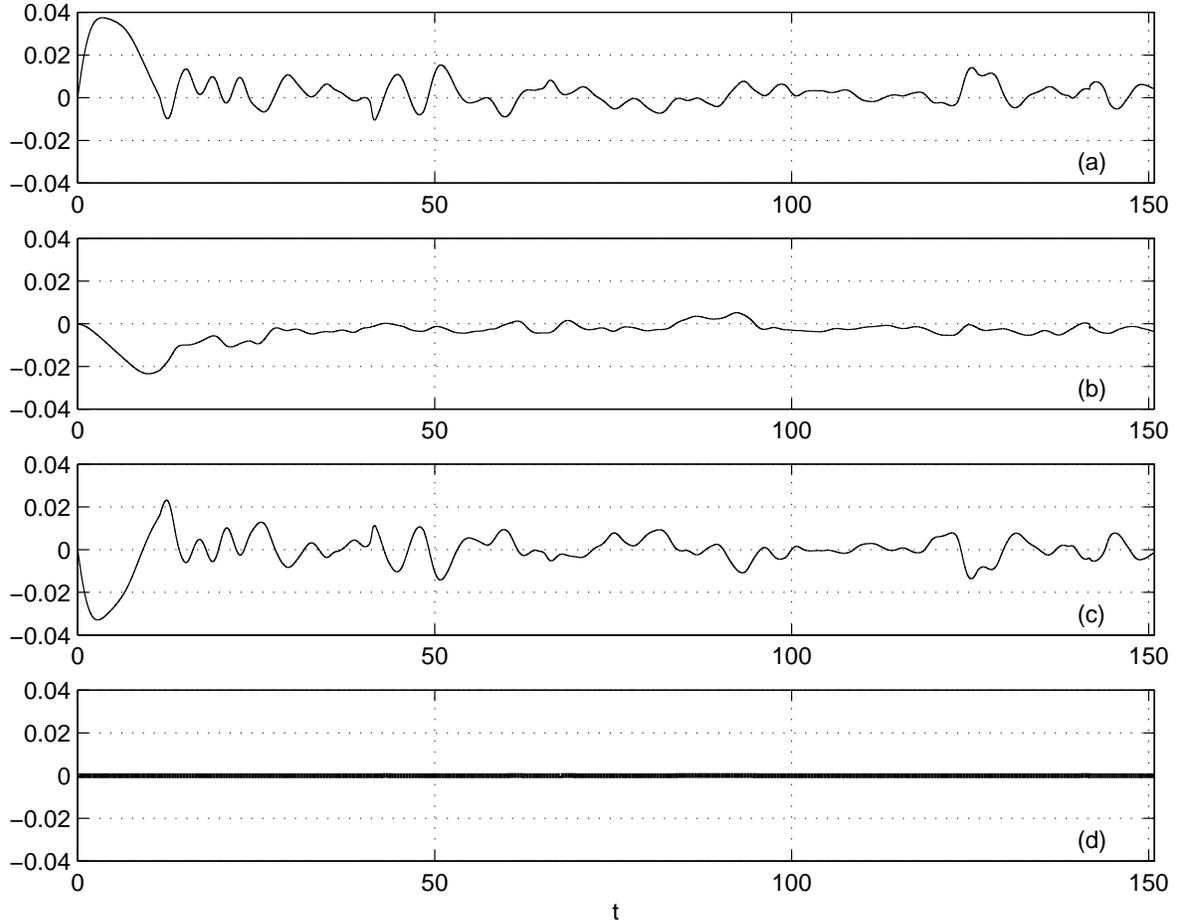}
\caption{Temporal evolution of the following quantities: (a) $2 \langle \v
\times \b \rangle  \cdot \OB$; (b) $2\la \langle \b \cdot \nabla \times \b \rangle $; (c)
$\part_t  \langle \a \cdot \b \rangle $; (d) sum of above three quantities. Data are
collected from Run I. }
\end{figure}

\begin{figure}
\plotone{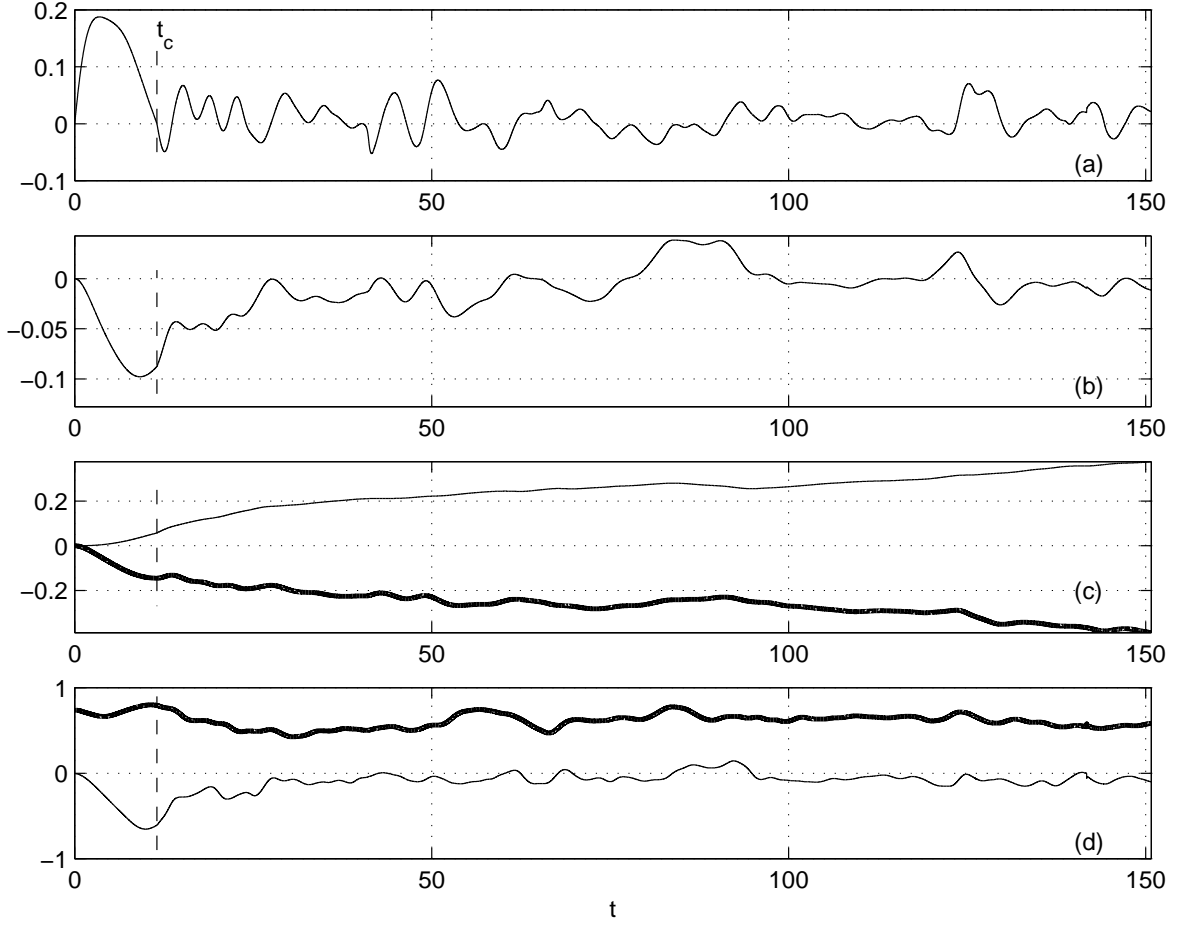}
\caption{Panel (a): $\alpha(t) = \frac{ \langle \v \times \b \rangle  \cdot
\OB}{{\overline B}^2}$. Panel (b): magnetic helicity density $ \langle \a
\cdot \b \rangle $. Panel (c): $-\int_0^{t}2\alpha(\tau) {\overline
B}^2 d\tau$(thick solid line) and
$-2\int_0^{t}\eta\langle\b\cdot\nabla\times\b\rangle$(thin solid
line). Panel (d): negative kinetic helicity density, $- \langle \v \cdot \nabla \times \v
\rangle $(thick solid line); current helicity density, $ \langle \b
\cdot \nabla \times \b \rangle $(thin solid line). $t_c \sim 11$. Data are collected from Run I. }
\end{figure}

\begin{figure}
\plotone{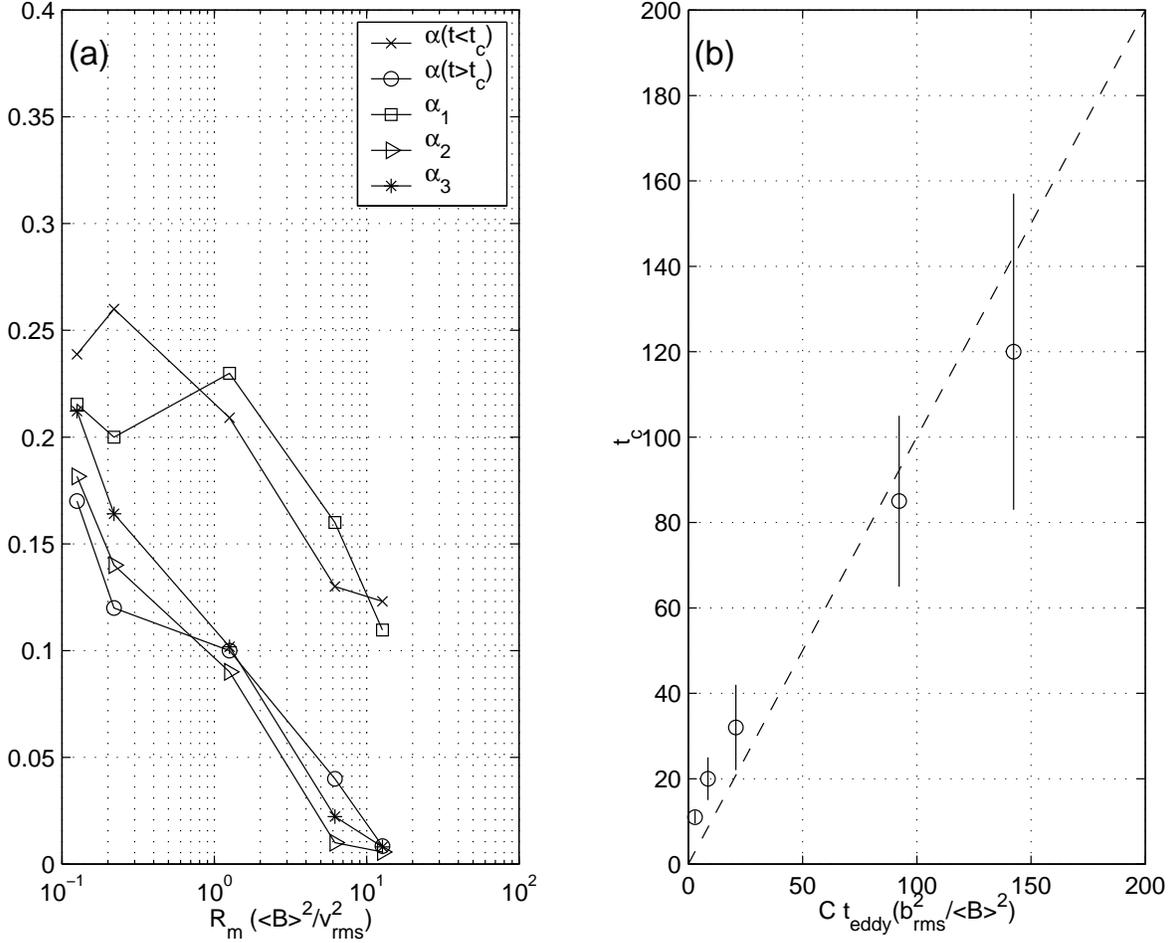}
\caption{Panel (a): The measured $\alpha(t<t_c)$, $\alpha(t>t_c)$, and
estimates of $\alpha$ based on different theoretical models (see text for
details). Panel (b): the critical time $t_c$ measured from the
simulation vs. the estimate based on our model (see section 1 of main
text). The correction factors to $t_c$ are estimated by reading the
spectrum of magnetic helicity and calculating the ratio
$v_{rms}/\alpha_{FBC}$. The correction factors are $C=1.5, 2, 1, 2.5$
and $3$ for Runs I to V. Error bars shown in panel (b) are the standard
deviation of our calculations for $t_c$. For the statistical standard
deviation calculations of other quantities in this figure, the reader
is referred to Table I for more details.}
\end{figure}

\end{document}